\documentclass{nature}
\usepackage{graphicx}
\usepackage{amsmath}
\usepackage{amssymb}
\usepackage{bm}
\usepackage{epstopdf}
\usepackage{epsfig}
\usepackage{color}
\usepackage{caption}

\title{Possible Electric-Field-Induced Superconducting States in Doped Silicene}

\author{Li-Da Zhang, Fan Yang \& Yugui Yao}

\begin{document}

\maketitle

\begin{affiliations}
\item[] School of Physics, Beijing Institute of Technology, Beijing 100081, China
\end{affiliations}

\begin{description}
\item[]
Correspondence and requests for materials should be addressed to F.Y. (yangfan\_blg@bit.edu.cn) or Y.Y. (ygyao@bit.edu.cn)
\end{description}

\begin{abstract}
Silicene has been synthesized recently, with experimental evidence showing possible superconductivity in the doped case. The noncoplanar low-buckled structure of this material inspires us to study the pairing symmetry of the doped system under a perpendicular external electric field. Our study reveals that the electric field induces an interesting quantum phase transition from the singlet chiral $d+id'$-wave superconducting phase to the triplet $f$-wave one. The emergence of the $f$-wave pairing results from the sublattice-symmetry-breaking caused by the electric field and the ferromagnetic-like intra-sublattice spin correlations at low dopings. Due to the enhanced density of states, the superconducting critical temperature of the system is enhanced by the electric field remarkably. Furthermore, we design a particular dc SQUID experiment to detect the quantum phase transition predicted here. Our results, if confirmed, will inject a new vitality to the familiar Si-based industry through adopting doped silicene as a tunable platform to study different types of exotic unconventional superconductivities.
\end{abstract}

The past few decades have witnessed the glorious development of unconventional superconductivity (SC) \cite{usc}. While most of the so far confirmed unconventional superconductors are singlet pairing including, say, the cuprates with $d$-wave pairing and the iron-pnictide with $s\pm$ pairing \cite{swave1,swave2}, no triplet superconductor has been confirmed except for the Sr$_2$RuO$_4$ system which shows evidence for possible $p+ip'$ pairing \cite{pip}. Recently, however, triplet pairing is catching more and more attentions partly due to its possible connection with topological SC \cite{tsc1,tsc2,tsc3,tsc4}, which has inspired great enthusiasm in searching for triplet superconductors. Here we predict a tunable quantum phase transition (QPT) from the singlet $d+id'$ superconducting state to the triplet $f$-wave one in doped silicene via applying a perpendicular external electric field on the system. Although both pairing symmetries break the time reversal symmetry, they are quite different. While the former shows chiral complex gap structures which is a hot topic recently \cite{did1,did2,did3,did4,did5,did6,did7}, the latter has a real pairing gap which changes sign with every $60^{o}$ rotation. The $f$-wave pairing has been proposed in the study of the quasi-1D organic system \cite{Q1D1}, the triangular lattice system for Na$_x$CoO$_2$ \cite{triangular}, and the spinless fermion system on the honeycomb optical lattice \cite{honeycomb}. However, unambiguously confirmed experimental evidence for it is still lack now. If the $f$-wave SC predicted here is confirmed experimentally, it will be the first time to realize such an intriguing high-angular-momentum pairing state.

Silicene, a single atomic layer of Si forming a 2D honeycomb lattice, can be regarded as the Si-based counterpart of graphene. It has attracted a lot of research interest since being successfully synthesized recently \cite{syn1,syn2,syn3,syn4,syn5}. Similar lattice structures between graphene and silicene bring them similar band structures with linear dispersion near the Fermi surface (FS), which further lead to similar physical properties between them. The most important structural difference between the two layered systems lies in that, while the graphene layer forms a regular flat plane, the silicene layer instead takes the form of noncoplanar low-buckled (LB) structure, with the sublattices A and B forming two separate planes. The LB structure of silicene causes many fascinating consequences, such as the enhanced quantum spin Hall effect \cite{qshe1,qshe2}, the quantum anomalous Hall effect and valley polarized quantum Hall effect in external electric field \cite{qahe1,qahe2,qahe3,qahe4}, and possible $d+id'$ chiral SC in bilayer silicene \cite{bilayer}. On the other hand, this LB structure provides the possibility to use gated silicene as a tunable source of the perfectly spin-polarized electrons \cite{gatesili}. Interestingly, a recent tunnelling experiment reported electronic gap in doped silicene \cite{gap}, probably caused by SC, which has attracted a lot of research interests \cite{interest1,interest2}.

In this paper, we report our study of the pairing symmetry of doped silicene under a perpendicular external electric field, which differentiates the on-site energies of the two sublattices of this LB system. Based on the random phase approximation (RPA) to the Hubbard model of the system, we reveal a tunable QPT from the singlet $d+id'$ chiral SC under weak field to the triplet $f$-wave one under strong field, without the necessity of long-range Coulomb interaction \cite{longrange1,longrange2,longrange3}. The physics behind this interesting QPT lies in that, under the strong sublattice-symmetry-breaking electric field, only one sublattice is left as the low energy subspace, and the ferromagnetic-like intra-sublattice spin correlations favor triplet $f$-wave pairing in this subspace. Due to the enhanced density of states (DOS) near the Fermi level, the superconducting critical temperature of the system is enhanced by the applied electric field remarkably. To detect the QPT predicted here, we further design a particular experiment with a phase-sensitive dc SQUID, which can distinguish between the $d+id'$ and $f$-wave pairings. Our study will open up a new era to utilize the familiar Si-based material as a tunable platform to study the competition and QPT among different types of exotic unconventional SCs.

\section*{Results}
\subsection{Model.}
The LB honeycomb lattice of silicene is shown in Figs. 1(a) and 1(b). Since the two sublattices A and B lie in two parallel planes separated from each other, an applied perpendicular electric field induces an on-site energy difference $\Delta$ between them. We adopt the following Hubbard model as an appropriate start point to study the low energy physics of the system:
\begin{align}\label{H}
H=-t\sum_{\langle ij\rangle\sigma}(c_{i\sigma}^{\dagger}c_{j\sigma}+h.c.)
+\frac{\Delta}{2}\sum_{i\sigma}(-1)^{p_i}c_{i\sigma}^{\dagger}c_{i\sigma}
+U\sum_in_{i\uparrow}n_{i\downarrow}.
\end{align}
Here the $t$-term (with $t=1.12$eV \cite{qshe2}) describes the nearest-neighbor (NN) hoppings, and $(-1)^{p_i}=+1$ ($-1$) for sublattice A (B). The Hubbard interaction strength $U=2$eV is adopted in the following, which is near the value obtained from the first principle calculations \cite{Ueff}. The next-nearest-neighbor (NNN) hoppings are not included here because they are qualitatively unimportant. The resulting particle-hole symmetry enables us to focus the following discussions only on the electron-doped case.

The band structure of the system with the dispersion $\varepsilon^{\pm}_{\bm{k}}
=\pm\sqrt{\left(\frac{\Delta}{2}\right)^2+t^2\left|1+e^{ik_x}+e^{-ik_y}\right|^2}$
is shown in Fig. 1(c). Clearly, the applied field induces a band gap $\Delta$ near the K-point and the nearby low energy band structure is flattened, although it does not modify the shape of the FS for a given doping concentration. Note that the electric field breaks the sublattice symmetry of the system. In particular, the on-site energy of sublattice A, i.e., $V_A=\Delta/2$, is closer to the Fermi level shown for the electron-doped case, and consequently this sublattice will dominate the low energy physics of the system. This effect turns out to be very important for our following discussions.

\subsection{Susceptibilities.}
According to the standard RPA approach \cite{bilayer,rpa1,rpa2,rpa3}, we define the free susceptibility ($U=0$) of the model as
\begin{align}\label{chi0t}
\chi^{(0)pq}_{st}(\bm{k},\tau)\equiv
\frac{1}{N}\sum_{\bm{k}_1\bm{k}_2}\left\langle
T_{\tau}c_{p}^{\dagger}(\bm{k}_1,\tau)
c_{q}(\bm{k}_1+\bm{k},\tau)
c_{s}^{\dagger}(\bm{k}_2+\bm{k},0)
c_{t}(\bm{k}_2,0)\right\rangle,
\end{align}
where $p,q,s,t=1,2$ is the sublattice index. The Hermitian static susceptibility matrix is defined as $\chi^{(0)}_{p,s}(\bm{k})\equiv\chi^{(0)pp}_{ss}(\bm{k},i\omega_n=0)$. The largest eigenvalue of this matrix represents the static susceptibility of the system in the strongest channel, and the corresponding eigenvector determines the pattern of the applied detecting field in that channel. In addition, the eigenvector also describes the pattern of the dominant intrinsic spin fluctuations in the system.

In Fig. 2(a)-2(c), we show the $\bm{k}$-space distributions of the zero temperature static susceptibility of the system in the strongest channel mentioned above. From Fig. 2(a) to 2(c), one can clearly observe the doping evolution of the static susceptibility. In particular, when the doping increases gradually from zero to the Van Hove (VH) doping $x=1/4$, the momenta of the maximum susceptibility evolves from the $\Gamma$-point (2(a)) first to a small circle around it (2(b)), and finally to the M-points (2(c)). Such an doping evolution of the susceptibility originates from the evolution of the FS which grows gradually from the K-points (the Dirac-points) first to small pockets around them, and finally to the connected large FS with perfect nesting at the VH singularity. From 2(a) to 2(c), one verifies that the dominant spin fluctuations on each sublattice of the system changes gradually with doping from ferromagnetic-like to antiferromagnetic-like. In Fig. 2(d), a typical pattern of the dominant spin fluctuations of the system is shown at $10\%$ doping for $\Delta=1$eV. Most prominently in Fig. 2(d), the magnetic moments are asymmetrically distributed between the two sublattices, with the magnitude of the moment on sublattice A obviously larger than that on sublattice B. This is consistent with the above band structure analysis which suggests that sublattice A will dominate the low energy physics of the system. Our calculation reveals that such asymmetry is enhanced by both electric-field and doping.

When the Hubbard interaction is turned on, the charge ($c$) or spin ($s$) susceptibility is given by RPA as
\begin{align}\label{chics}
\chi^{(c(s))}(\bm{k},i\omega_n)=
\left[I\pm\chi^{(0)}(\bm{k},i\omega_n)(U)\right]^{-1}
\chi^{(0)}(\bm{k},i\omega_n),
\end{align}
where $(U)$ is a $4\times4$ matrix, whose only two nonzero elements are $(U)^{\mu\mu}_{\mu\mu}=U$ ($\mu=1,2$) \cite{bilayer}. Clearly, the repulsive Hubbard interaction suppresses $\chi^{(c)}$ and enhances $\chi^{(s)}$. When the interaction strength $U$ is greater than a critical value $U_c$, the spin susceptibility $\chi^{(s)}$ of the model diverges, which implies the instability towards the long-range spin-density-wave (SDW) order. Below the critical interaction strength $U_c$, the spin fluctuations take the main role in mediating the Cooper pairing.

\subsection{Pairing symmetries.}
Through exchanging the renormalized susceptibilities (\ref{chics}), one obtains the effective interaction vertex $V^{\alpha\beta}(\bm{k},\bm{k}')$ \cite{rpa3} between two Cooper pairs on the FS, which is then plugged into the following linearized gap equation near the superconducting critical temperature $T_c$ \cite{rpa3}:
\begin{align}\label{gapeq}
-\frac{1}{(2\pi)^2}\sum_{\beta}\oint_{FS}dk'_{\parallel}
\frac{V^{\alpha\beta}(\bm{k},\bm{k}')}{v^{\beta}_F(\bm{k}')}
\Delta_\beta(\bm{k}')=\lambda\Delta_\alpha(\bm{k}).
\end{align}
Here the integration is along various FS patches labelled by $\alpha$ or $\beta$, $v^{\beta}_F(\bm{k}')$ is the Fermi velocity and $k'_\parallel$ is the tangential component of $\bm{k}'$ along the FS. Solving Eq. (\ref{gapeq}) as an eigenvalue problem, one obtains the largest eigenvalue $\lambda$, which determines $T_c$ via $T_c\approx te^{-1/\lambda}$, and the corresponding eigenvector $\Delta_\alpha(\bm{k})$ as the leading gap form factor.

The phase diagram on the $x$-$\Delta$ plane obtained by our RPA calculations is shown in Fig. 3(a). Clearly, except for the SDW phase near the VH doping $x=1/4$, the singlet chiral $d_{x^2-y^2}+id_{xy}$ ($d+id'$) and triplet nodeless $f$-wave pairings beat other instabilities and serve as the leading instability in different regimes of the phase diagram. The gap function of the $d_{xy}$ symmetry (for doping $x=0.15$ and $\Delta=0$) shown on the FS in Fig. 3(b) is antisymmetric about the $x$ and $y$ axes, and that of its degenerate partner $d_{x^2-y^2}$ (not shown) is symmetric about these axes. Their mixing in the form of $d+id'$ minimizes the ground state energy. The main part of the $d+id'$ pairing in real space is distributed on NN-bonds as shown in Fig. 3(d). The gap function of the $f$-wave pairing (for doping $x=0.15$ and $\Delta=1$eV) is shown in Fig. 3(c). The main part of the $f$-wave pairing in real space is distributed on NNN-bonds as shown in Fig. 3(e). Clearly, the gap function of this time-reversal-breaking $f$-wave pairing changes sign with every $60^o$ rotation either in $\bm{k}$-space or in real space.

As shown in Fig. 3(a), the leading instability for $\Delta=0$ is $d+id'$ pairing at low dopings and SDW order near the VH doping. This result is consistent with previous calculations on the graphene system \cite{did1,did2,did3,did4,did5,did6}, which shares the same Hubbard model as here. In Fig. 3(f), we show the doping dependence of the SDW critical interaction $U_c$. Clearly, when the doping $x\gtrsim0.2$, $U_c$ is less than $U=2$eV, which accounts for the emerging of the SDW state in Fig. 3(a). The most interesting and important discovery here is that the triplet $f$-wave pairing, which is mediated by the NNN-bond ferromagnetic spin fluctuations, wins over the $d+id'$ pairing and rises as the leading pairing symmetry when sufficiently strong electric-field is applied on the system at low dopings. Physically, such an electric-field-induced QPT originates from the sublattice-symmetry-breaking effect mentioned before, which has selected the sublattice A as the low energy subspace of the system. All the relevant low energy physics, including the pattern of the spin fluctuations shown in Fig. 2(d), the effective pairing interaction mediated by these spin fluctuations, and the pairing itself take place mainly in this subspace in strong electric field. Consequently, at low dopings, the inter-sublattice pairing $d+id'$ symmetry shown in Fig. 3(d) has to give way to the intra-sublattice pairing $f$-wave symmetry shown in Fig. 3(e).

The superconducting critical temperature $T_c\approx te^{-1/\lambda}$ of the $f$-wave pairing state is controllable, and can be remarkably enhanced by the applied electric field. As shown in Fig. 3(g) for doping $x=0.15$, the eigenvalue $\lambda$ of the $f$-wave pairing can be enhanced to about $0.18$ when $\Delta$ is tuned to $1$eV. Although the RPA approach tends to overestimate $T_c$, we still have space to enhance the $T_c$ of the system to the experimentally accessible range by tuning the doping level closer to the VH doping or increasing the electric field. Physically, such enhancement is attributed to the increase of the DOS near the Fermi level (see the inset of Fig. 3(g)) caused by the flattening of the band structure under the electric field (see Fig. 1(c)). The increase of the DOS not only enhances the number of Cooper pairs near the FS, but also enhances the pairing interaction. The $T_c$ of the $d+id'$ pairing can also be enhanced by sufficiently strong electric field due to this DOS enhancement, but it is lower than that of the $f$-wave pairing as shown in Fig. 3(g). For a vanishingly small $U$, the $f$-wave SC has been proposed even in the electric-field-free system \cite{smallu}, but with extremely low $T_c$.

\subsection{Detecting the QPT.}
The QPT from $d+id'$ to $f$-wave pairings predicted here can be detected by the dc SQUID, a phase-sensitive device which has been adopted in determining the pairing symmetries of such superconducting systems as cuprates \cite{squid} and Sr$_2$RuO$_4$ \cite{josephson}. Our basic scheme is shown in Figs. 4(a) and 4(d), where a slice of silicene is fabricated into a hexagonal shape, allowing the relative phase among different directions in the system to be detected. Two superconductor-normal metal-superconductor (SNS) Josephson tunneling junctions are formed on the opposite (4(a)) or adjacent (4(d)) edges of the hexagon, which are connected by a loop of a conventional $s$-wave superconductor, forming a bimetallic ring with a magnetic flux $\Phi$ threading through the loop.

As a result of the interference between the two branches of Josephson supercurrent, the maximum total supercurrent (the critical current) $I_c$ in the circuit modulates with $\Phi$ according to
\begin{align}
I_c(\Phi)=2I_{0}\left|\cos\left(\pi\frac{\Phi}{\Phi_0}
+\frac{\delta_{ab}}{2}\right)\right|.
\end{align}
Here $I_0$ is the critical current of one Josephson junction, $\Phi_{0}=h/2e$ is the basic flux quantum, and $\delta_{ab}$ is the intrinsic phase shift inside the silicene system between pairs tunneling into the system in two directions a and b. In configuration 4(a), $\delta_{ab}$ is equal to $0$ ($\pi$) for singlet pairing including the $d+id'$ one (triplet pairing including the $f$-wave one), and in configuration 4(d), it is $2\pi/3$ ($\pi$) for the $d+id'$ ($f$) symmetry. Consequently, the $I_c\sim\Phi$ curves for $d+id'$ and $f$-wave pairings in configurations 4(a) and 4(d) show different patterns (Figs. 4(b), 4(c), 4(e), and 4(f)). Most prominently, the $f$-wave pairing symmetry is characterized by the minima of $I_c$ (which can be nonzero when the self-inductance of the loop is not negligible) at zero flux for both configurations 4(a) and 4(d). For the $d+id'$ symmetry at zero flux, while the critical current $I_c$ is a maximum in configuration 4(a), it is not an extreme point in 4(d). Thus, by observing the modulation of the SQUID response vs applied magnetic flux, one can distinguish between the two pairing symmetries and hence detect the QPT.

\section*{Discussion}
The non-monotonic doping dependence of the critical electric field in Fig. 3(a) originates from the competition between two opposite effects. On the one hand, the sublattice-symmetry-breaking effect, which favors the $f$-wave pairing, is enhanced by doping since the Fermi level sits in the middle of $V_A$ and $V_B$ at zero doping (see Fig. 1(c)). On the other hand, as shown in Fig. 2(a)-2(c), the intra-sublattice spin correlations in the system evolves from ferromagnetic-like at low dopings, which favors triplet $f$-wave pairing, to antiferromagnetic-like near the VH doping, which favors singlet $d+id'$ pairing.

Besides the nodeless $f$-wave symmetry predicted here, there is also another nodal $f'$-wave symmetry which is mainly composed of intra-sublattice pairings with a bond length of three times of the lattice constant. This $f'$-wave symmetry is not favored here, for its gap nodes along the $\Gamma$-K lines do not avoid the FS. When a sufficiently strong Kane-Mele spin-orbit coupling (SOC) \cite{soc1,soc2} is added into model (\ref{H}), the Fermi pockets of the system would shift from near the K-points to near the M-points, and the $f'$-wave symmetry would be favored, for its gap nodes avoid the FS. In a system with a medium SOC, the two $f$-wave pairings could be nearly degenerate, and their mixing in the form of $f+if'$ might be realized to minimize the energy. As a result of its nontrivial topological property, this intriguing triplet chiral superconducting state can harbor the Majorana zero-mode at its boundary \cite{tsc2,maj1,maj2,maj3}, which is useful in the topological quantum computation. Although the SOC in the present silicene system is too small \cite{qshe1,qshe2} to favor the $f+if'$ pairing, we can expect the system with a LB honeycomb lattice similar to silicene and a medium SOC, such as BiH \cite{bih1,bih2}, could serve as the platform for this novel high-angular-momentum chiral pairing state. We leave this subject for future studies.

As a result of the rapid development of modern experimental techniques, the electric field strong up to $0.3$V/{\AA} has already been applied to the research works of materials \cite{field}, which has approached what we need here. Besides, the monolayer of silicene has been synthesized on top of the substrates of Ag, ZrB$_2$, Ir, etc \cite{syn1,syn2,syn3,syn4,syn5}. Due to the low-buckled structure of silicene, the distance between the sublattice A and the substrate is different from that between the sublattice B and the substrate. As a result, the interaction between the silicene monolayer and the substrate provides an effective electric field, which causes the sublattice-symmetry-breaking and the difference between the on-site energies of the two sublattices \cite{break}. Since such an effective electric field is an internal electric field of chemical origin, it can generally be as strong as what we need here. Therefore, our proposal of the $f$-wave pairing is feasible in practice.

In conclusion, we have systematically studied the pairing phases of doped silicene in a perpendicular external electric field. The results of our RPA study predict that with the enhancement of the electric field, the system will experience a QPT from singlet $d+id'$ superconducting state to triplet $f$-wave one, and the superconducting critical temperature of the system will be enhanced due to the increase of the DOS. Our model needs neither long-range Coulomb interaction nor the situation of vanishingly small Hubbard-$U$, and thus is more realizable than other proposed ones.

\section*{Methods}
\subsection{Susceptibilities.}
According to the standard RPA approach \cite{bilayer,rpa1,rpa2,rpa3} adopted in our study, we first define the free susceptibility ($U=0$) of the model (\ref{H}) as in Eq. (\ref{chi0t}). Direct calculation yields the following explicit expression of the free susceptibility,
\begin{align}\label{chi0w}
\chi^{(0)pq}_{st}(\bm{k},i\omega_n)=\frac{1}{N}
\sum_{\bm{k}'\alpha\beta}(\xi^{\alpha}_{\bm{k}'})_t
(\xi^{\alpha}_{\bm{k}'})_p^*(\xi^{\beta}_{\bm{k}'+\bm{k}})_q
(\xi^{\beta}_{\bm{k}'+\bm{k}})_s^*
\frac{n_F(\varepsilon^{\beta}_{\bm{k}'+\bm{k}})
-n_F(\varepsilon^{\alpha}_{\bm{k}'})}
{i\omega_n+\varepsilon^{\alpha}_{\bm{k}'}
-\varepsilon^{\beta}_{\bm{k}'+\bm{k}}},
\end{align}
where $\varepsilon^{\alpha}_{\bm{k}}$ and $\xi^{\alpha}_{\bm{k}}$ are the $\alpha$-th eigenvalue and eigenvector of the single particle Hamiltonian of the system respectively, and $n_F$ is the Fermi-Dirac distribution function.

When the interaction is turned on, we define the charge ($c$) and spin ($s$) susceptibilities as
\begin{align}\label{chics0}
\chi^{(c)pq}_{st}(\bm{k},\tau)\equiv&\frac{1}{2N}
\sum_{\substack{\bm{k}_1\bm{k}_2\\\sigma_1\sigma_2}}\left\langle T_{\tau}c^{\dagger}_{p\sigma_1}(\bm{k}_1,\tau)
c_{q\sigma_1}(\bm{k}_1+\bm{k},\tau)
c^{\dagger}_{s\sigma_2}(\bm{k}_2+\bm{k},0)
c_{t\sigma_2}(\bm{k}_2,0)\right\rangle,         \\
\chi^{(s)pq}_{st}(\bm{k},\tau)\equiv&\frac{1}{2N}
\sum_{\substack{\bm{k}_1\bm{k}_2\\\sigma_1\sigma_2}}
\sigma_1\sigma_2\left\langle T_{\tau}
c^{\dagger}_{p\sigma_1}(\bm{k}_1,\tau)
c_{q\sigma_1}(\bm{k}_1+\bm{k},\tau)
c^{\dagger}_{s\sigma_2}(\bm{k}_2+\bm{k},0)
c_{t\sigma_2}(\bm{k}_2,0)\right\rangle.
\end{align}
For $U=0$, we have $\chi^{(c)}=\chi^{(s)}=\chi^{(0)}$. In the RPA level, $\chi^{(c(s))}$ is given by Eq. (\ref{chics}).

\subsection{Effective interaction and gap equation.}
Consider the scattering of a Cooper pair from the state $(\bm{k}',-\bm{k}')$ in the $\beta$-th ($\beta=1,2$) band to the state $(\bm{k},-\bm{k})$ in the $\alpha$-th ($\alpha=1,2$) band. This scattering process can be described by the following effective interaction,
\begin{align}\label{veff}
V_{eff}=&\sum_{\bm{k}\bm{k}'\alpha\beta}
V^{\alpha\beta}(\bm{k},\bm{k}')c^{\dagger}_{\alpha}(\bm{k})
c^{\dagger}_{\alpha}(-\bm{k})c_{\beta}(-\bm{k}')c_{\beta}(\bm{k}').
\end{align}
Here the projective interaction vertex $V^{\alpha\beta}(\bm{k},\bm{k}')$ is given by the effective vertex $\Gamma^{pq}_{st}(\bm{k},\bm{k}')$ through
\begin{align}\label{vab}
V^{\alpha\beta}(\bm{k},\bm{k}')=Re\sum_{\substack{\bm{k}\bm{k}'\\pqst}}
\Gamma^{pq}_{st}(\bm{k},\bm{k}')(\xi^{\alpha}_{\bm{k}})_p^*
(\xi^{\alpha}_{-\bm{k}})_q^*(\xi^{\beta}_{-\bm{k}'})_s
(\xi^{\beta}_{\bm{k}'})_t,
\end{align}
and $\Gamma^{pq}_{st}(\bm{k},\bm{k}')$ itself, in the singlet channel, reads
\begin{align}\label{gams}
\Gamma^{pq}_{st}(\bm{k},\bm{k}')=(U)^{pt}_{qs}
&+\frac{1}{4}\left\{U\left[3\chi^{(s)}(\bm{k}-\bm{k}')
-\chi^{(c)}(\bm{k}-\bm{k}')\right]U\right\}^{pt}_{qs}         \nonumber\\
&+\frac{1}{4}\left\{U\left[3\chi^{(s)}(\bm{k}+\bm{k}')
-\chi^{(c)}(\bm{k}+\bm{k}')\right]U\right\}^{ps}_{qt},
\end{align}
and, in the triplet channel, reads
\begin{align}\label{gamt}
\Gamma^{pq}_{st}(\bm{k},\bm{k}')=
&-\frac{1}{4}\left\{U\left[\chi^{(s)}(\bm{k}-\bm{k}')
+\chi^{(c)}(\bm{k}-\bm{k}')\right]U\right\}^{pt}_{qs}         \nonumber\\
&+\frac{1}{4}\left\{U\left[\chi^{(s)}(\bm{k}+\bm{k}')
+\chi^{(c)}(\bm{k}+\bm{k}')\right]U\right\}^{ps}_{qt}.
\end{align}
The mean-field decoupling of the effective interaction (\ref{veff}) in the Cooper channel gives rise to the self-consistent gap equation. Near the critical temperature $T_c$, this equation is linearized as Eq. (\ref{gapeq}), solving which one obtains the leading pairing symmetries and the corresponding $T_c$.

If the leading pairing symmetries include two degenerate doublets such as $d_{x^2-y^2}$ and $d_{xy}$, we can express the gap function of the system as
\begin{align}\label{did}
\Delta_\alpha(\bm{k})=C_1d^\alpha_{x^2-y^2}(\bm{k})
+(C_2+iC_3)d^\alpha_{xy}(\bm{k}).
\end{align}
Here $d^\alpha_{x^2-y^2}(\bm{k})$ and $d^\alpha_{xy}(\bm{k})$ represent the normalized gap functions of corresponding symmetries. Then the mixing coefficients $C_1$, $C_2$, and $C_3$ are determined by the minimization of the total mean-field energy with the constraint of the average electron number in the superconducting state.

\begin{addendum}
\item[Acknowledgements]
We are grateful to Hong Yao, Qiang-Hua Wang, Yi Zhou and Yu-Zhong Zhang for stimulating discussions. The work is supported by the MOST Project of China (Grants Nos. 2014CB920903, 2011CBA00100), the NSF of China (Grant Nos. 11174337, 11225418, 11274041, 11334012), and the Specialized Research Fund for the Doctoral Program of Higher Education of China (Grants Nos. 20121101110046, 20121101120046). F.Y. is supported by the NCET program under Grant No. NCET-12-0038.

\item[Author Contributions]
Y.Y. and F.Y. conceived the idea of searching the pairing symmetry in doped silicene subject to an applied electric field. L.D.Z. carried out the RPA calculations. All authors analyzed the data and wrote the manuscript.

\item[Competing Interests]
The authors declare that they have no competing financial interests.
\end{addendum}

\begin{figure*}
\centering
\epsfig{file=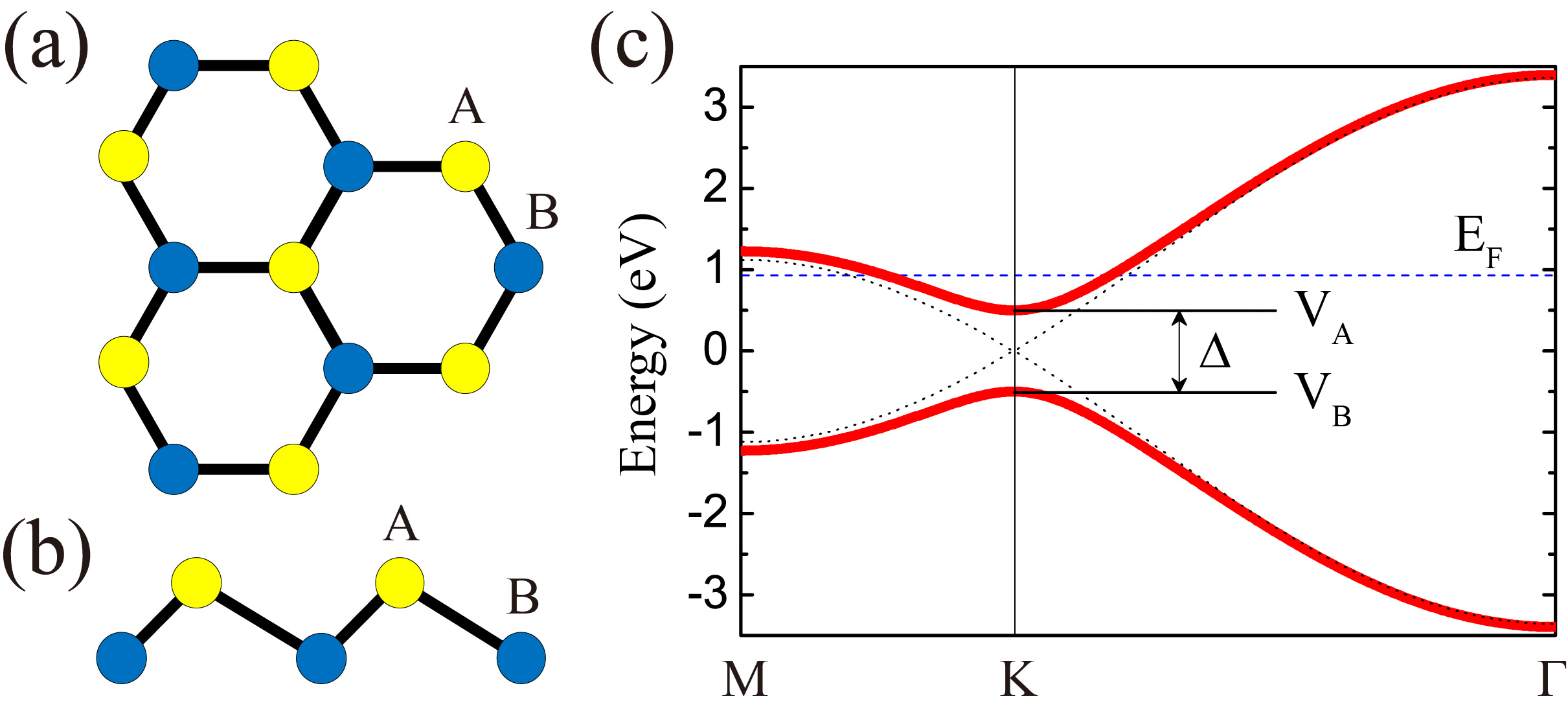,width=16cm}
\caption{{\bf Lattice and band structures of silicene.} (a) The top view and (b) the side view of the noncoplanar low-buckled lattice structure of silicene with the yellow and blue dots denote sublattices A and B respectively. (c) The band structures for $\Delta=0$ (black dotted lines) and $\Delta=1$eV (red solid lines) with the Fermi level for electron-doping $x=0.1$ and $\Delta=1$eV.}
\end{figure*}

\begin{figure*}
\centering
\epsfig{file=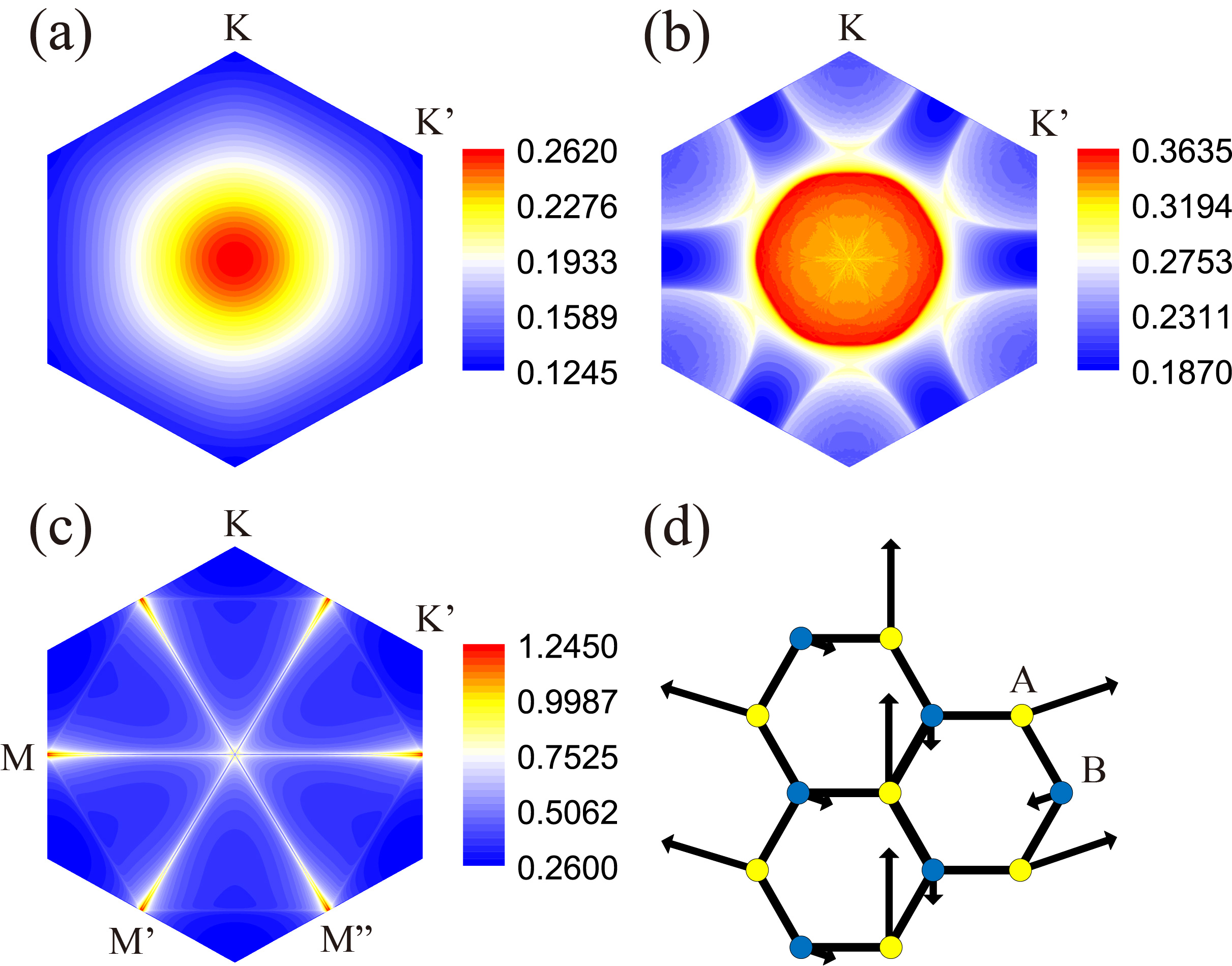,width=16cm}
\caption{{\bf Static susceptibility and the dominant spin fluctuations.} The $\bm{k}$-space distributions of the zero temperature static susceptibility for $\Delta=1$eV and different dopings: (a) $x=0$, (b) $x=0.1$, and (c) $x=0.25$. (d) A typical pattern of the dominant spin fluctuations for $x=0.1$ and $\Delta=1$eV.}
\end{figure*}

\begin{figure*}
\centering
\epsfig{file=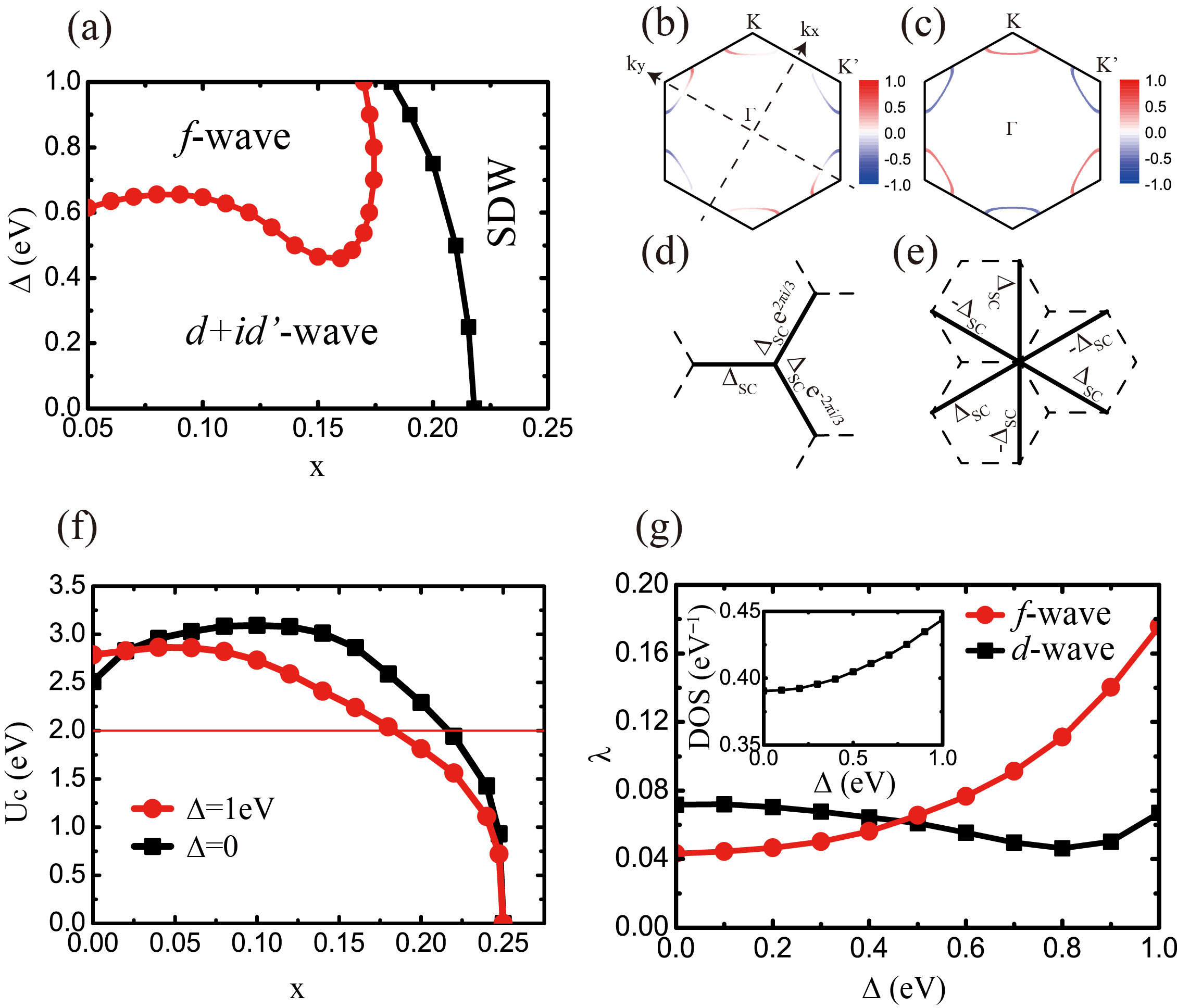,width=16cm}
\caption{{\bf Superconducting phase diagram, gap functions, critical interactions and  pairing eigenvalues.} (a) The superconducting phase diagram of the system on the $x$-$\Delta$ plane. Distributions of the gap functions on the FS for doping $x=0.15$: (b) $d_{x^2-y^2}$ symmetry for $\Delta=0$ and (c) $f$-wave symmetry for $\Delta=1$eV. The main parts of the real-space pairings for (d) $d+id'$ symmetry and (e) $f$-wave symmetry. (f) The doping dependence of the SDW critical interactions $U_c$ for $\Delta=0$ and $\Delta=1eV$. (g) The electric-field dependences of the largest eigenvalues $\lambda$ of $f$-wave and $d$-wave pairings for doping $x=0.15$. Inset: the electric-field dependence of DOS at the Fermi level for $x=0.15$.}
\end{figure*}

\begin{figure*}
\centering
\epsfig{file=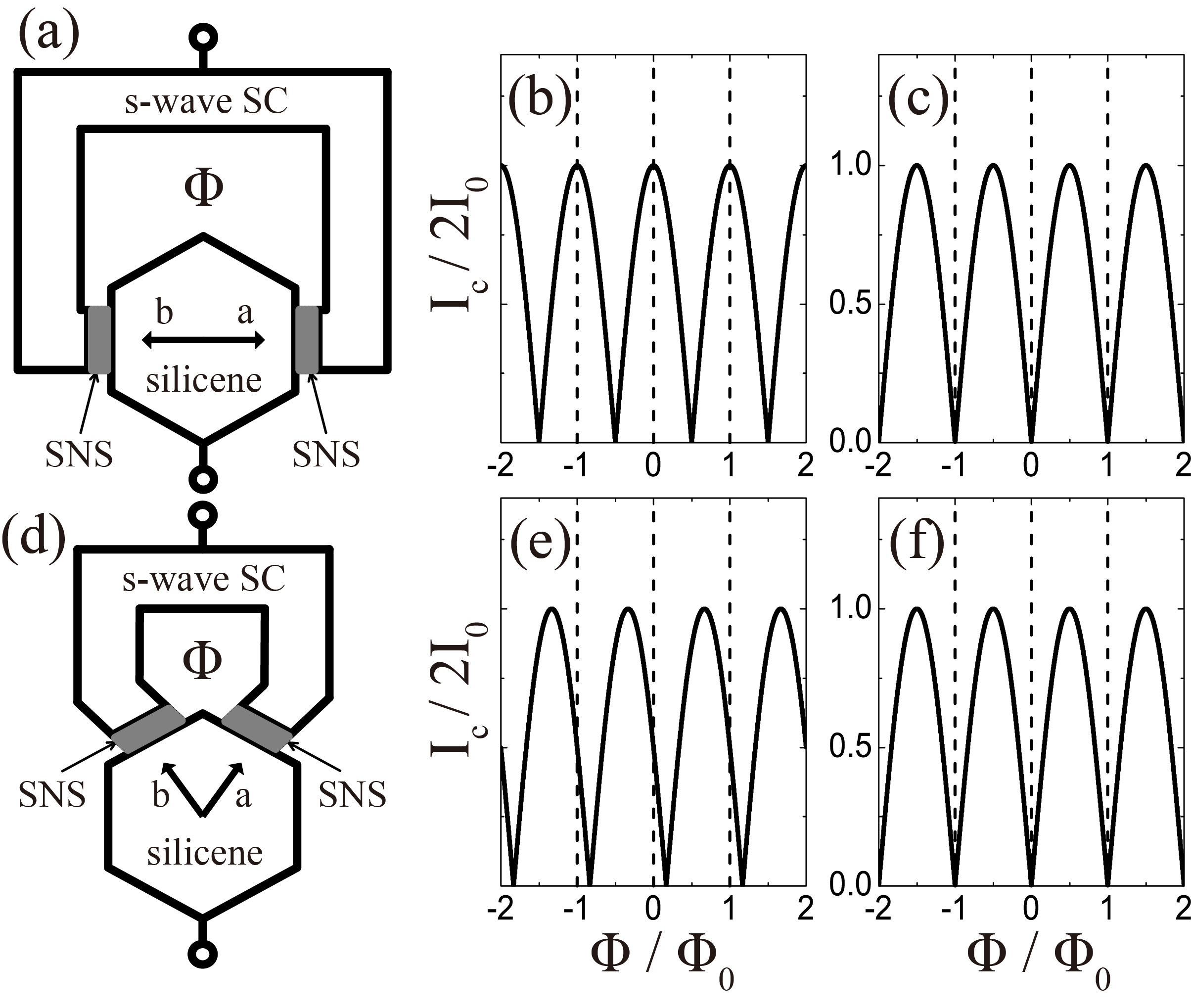,width=16cm}
\caption{{\bf Design of the SQUID experiment.} Configurations of the dc SQUID interferometer experiments used to determine the relative phase (a) between the opposite edges and (d) between the adjacent edges of the hexagon of silicene, where $\Phi$ represents the magnetic flux. The expected $\Phi$-dependences of the critical current $I_c$ for (b) singlet pairing including the $d+id'$ one and for (c) triplet pairing including the $f$-wave one corresponds to configuration (a). The expected results for (e) the $d+id'$ pairing and for (f) the $f$-wave pairing corresponds to configuration (d).}
\end{figure*}

\end{document}